\documentclass[12pt]{iopart}

\usepackage{graphicx}  

\begin{document}

\title[Strain and stacking faults in single nanowires using Bragg coherent diffraction]{Analysis of strain and stacking faults in single nanowires using Bragg coherent diffraction imaging}

\author{V Favre-Nicolin$^{1,2}$, F Mastropietro$^{1,3}$, J Eymery$^{1}$, D Camacho$^{1}$, Y M Niquet$^{1}$, B M Borg$^4$, M E Messing$^4$, L-E Wernersson$^4$, R E Algra$^{5,6,7}$, E P A M Bakkers$^6$, T H Metzger$^3$, R Harder$^8$ and I K Robinson$^{9,10}$}

\address{$^1$CEA, INAC, SP2M, 38054 Grenoble, France}
\address{$^2$Universit\'e Joseph Fourier, Grenoble, France}
\address{$^3$European Synchrotron Radiation Facility, 6 rue Jules Horowitz BP220, 38043 Grenoble Cedex, France}
\address{$^4$Department of Solid State Physics, Lund University, Box 118, S-221 00 Lund, Sweden}
\address{$^5$Materials innovation institute (M2i), 2628CD Delft, The Netherlands }
\address{$^6$Philips Research Laboratories, Eindhoven, High Tech Campus 11, 5656AE Eindhoven, The Netherlands}
\address{$^7$IMM, Solid State Chemistry, Radboud University Nijmegen, Heijendaalseweg 135,6525AJ Nijmegen, The Netherlands}
\address{$^8$Argonne National Laboratory, 9700 South Cass Avenue, Argonne, IL 60439, USA}
\address{$^9$London Centre for Nanotechnology, University College, 17–19 Gordon Street, London WC1H 0AH, UK}
\address{$^{10}$Diamond Light Source, Harwell Campus, Didcot, Oxfordshire OX11 0DE, UK}
\ead{Vincent.Favre-Nicolin@cea.fr}
\begin{abstract}
Coherent diffraction imaging (CDI) on Bragg reflections is a promising
technique for the study of three-dimensional (3D) composition and strain fields in
nano-structures, which can be recovered directly from the coherent diffraction data
recorded on single objects. In this article we report results obtained for single homogeneous and heterogeneous nanowires with a diameter smaller than 100 nm, for which we used CDI to retrieve information about deformation and faults existing in these wires. The article also discusses the influence of stacking faults, which can create artefacts during the reconstruction of the nanowire shape and deformation.

\end{abstract}

\pacs{61.46.Km, 62.23.Hj, 61.05.cp, 42.30.Rx}

\maketitle

\section{Introduction}
Coherent X-ray Diffraction Imaging (CXDI) has been developped for the past 10 years \cite{miao_extendingmethodology_1999,robinson_reconstruction_2001,pfeifer_three-dimensional_2006,livet_diffraction_2007}: thanks to the development of focusing optics \cite{snigirev_high_2008} such as compound refractive lenses (CRL) \cite{snigirev_focusing_1998}, Kirckpatrick-Baez mirrors (KB) \cite{kirkpatrick_formation_1948} and Fresnel-zone plates (FZP) \cite{baez_fresnel_1961,di_fabrizio_high-efficiency_1999}, it is now possible to collect the scattering of single objects with a size smaller than 100 nm \cite{schroer_coherent_2008}.

One goal of CXDI is to allow for the 3D reconstruction of single, non-crystalline objects such as biomolecules \cite{miao_approach_2001} or amorphous materials \cite{barty_three-dimensional_2008}: in this case the small-angle scattering is measured, optionally in 3D using a tomographic approach \cite{marchesini_coherent_2003}, and the electronic density of the sample can be reconstructed using an inverse Fourier transform combined with phase retrieval algorithms \cite{gerchberg_practical_1972,fienup_phase_1982,marchesini_x-ray_2003,wu_reconstruction_2005}.

In the case of crystalline materials, it is also possible to measure the scattered intensity around Bragg peaks \cite{robinson_reconstruction_2001,pfeifer_three-dimensional_2006,williams_three-dimensional_2003,stadler_coherent_2007}, a method which we will refer to as Coherent Bragg Imaging (CBI). CBI allows to recover the shape (the electronic density) of the objects, but this is not obviously the main interest of the method, as electron microscopes can much more easily give access to the shape of any free-standing nano-object, with a better resolution. The interest of CBI comes from the fact that the scattered intensity around a Bragg reflection is sensitive to the deformation of the crystal, at a resolution smaller than the d-spacing of the considered reflection. Moreover it is possible to study buried objects  in their normal environment (e.g. beneath a protecting shell) without sample preparation that could change their strain state. So far only a few examples of deformation reconstruction in nano-objects have been published using CBI \cite{pfeifer_three-dimensional_2006,harder_orientation_2007,robinson_coherent_2009}, generally with small deformations that were not due to a controlled epitaxial process.

Nanowires have recently been the focus of CBI experiments \cite{robinson_coherent_2009,chamard_evidence_2008,favre-nicolin_coherent-diffraction_2009,diaz_coherent_2009,leake_longitudinal_2009}, but so far no CBI on heterogenous nanostructures have been presented with a successful reconstruction of the shape and deformation field, although a few measurements on SiGe islands have been reported \cite{mocuta_beyondensemble_2008,scheler_probingelastic_2009,diaz_spatially_2009}. A possible explanation for the relative lack of publication in the field is the existence of structural faults: dislocations (inherent to heterogeneous structures grown by epitaxy) and stacking faults (related to the low stacking fault energy \cite{krishna_random_1994} often found for nanowires).

In this article we will discuss the importance of structural faults in coherent scattering experiments, and how they affect our ability to study heterogeneous nanostructures using CBI, in the particular case of nanowires.
\\

This paper is organized as follows: in section \ref{section2} we give a general presentation of coherent Bragg imaging in the case of strained nano-objects. Section \ref{section3} presents CBI results on homogeneous nanowires as well as perfect (simulated) heterogeneous nanowires. Section \ref{section4CBIFaultedNW} details how the presence of stacking faults affects CBI, and how it is possible to obtain quantitative results for strain mapping and a statistical analysis of the faults.
\\

The following crystallographic conventions will be used: in reciprocal space the scattering vector $\vec{k}$ will be given either in reciprocal lattice units (r.l.u.) $(hkl)$, or using its norm $\vec{k}=\frac{2\sin\vartheta}{\lambda}$. Displacement values (e.g. $u_x$) will be given in values normalized to the unit cell length (u.c.). As the nanowires considered here present several polymorphs, when ambiguous the reciprocal coordinates will be given a $ZB$ (zincblende) or $WZ$ (wurtzite) subscript, e.g. $(111)_{ZB}$ and $(002)_{WZ}$ which correspond to the same reflection with different reference lattices.
 
\section{Coherent Bragg Imaging on strained nano-objects: theory}\label{section2} 
In this section, we present the general theory of Bragg coherent X-ray diffraction applied to strained nano-objects, and particularly the different approximations that can be used to compute the scattering, depending on the type of structural model that is available.
\\

In the case of focused X-rays, the wavefield close to the focal point is not a strict plane-wave, due to the curvature of the incident wave which induces a variation in both the amplitude and the phase of the X-ray wavefield. However, if we assume that the scattering object is close enough to the focal point and is small compared to the size of the focal point, then a plane wave is a quite good approximation \cite{schroer_coherent_2008}.

For small objects ($<$200 nm) it is reasonable to evaluate the scattering within the kinematical approximation. However, in the case of larger objects with heavy materials, refraction effects must be taken into account \cite{harder_orientation_2007}. They will be ignored in this article (focused on nanowires with a diameter smaller than 100 nm) for the sake of simplicity.

Given these approximations the scattering of the X-rays is due to the interaction with the sample electrons, and can be written as the Fourier transform of the electronic density:
\begin{equation}
\tilde{A}(\vec{k})=\int_{V}\rho(\vec{r}) e^{2i\pi\vec{k}\cdot\vec{r}}=FT\left[\rho(\vec{r})\right]
\end{equation}
where $\tilde{A}$ is the complex scattered amplitude for the scattering vector $\vec{k}=\vec{k_f}-\vec{k_i}$, $\rho(\vec{r})$ is the electronic density in the sample, and $FT$ denotes the Fourier transform.

This formula is suitable for the analysis of CDI experiments in the (relatively) small angle regime, i.e. for small $\vec{k}$ values: indeed the real-space resolution at which the electronic density must be described is directly related to the extent of the Fourier transform in reciprocal space. For example in the case of a nanocube of silicon with a size of 100 unit cells (54.3 nm), in order to compute the scattering up to a (moderate) 0.1\ nm resolution, the size of the electronic density array would be $543^3\approx160\times10^6$ points.

This approach is unsuitable for the study of strained nano-objects using CBI, where the scattering will be collected far from the small angle regime, especially as the displacement from a perfect periodic lattice will often be a fraction of the unit cell's dimensions. In this case it is preferable to use an atomistic description of the crystal to compute its scattering, i.e.:
\begin{equation}\label{eqn:eqn2} 
\tilde{A}(\vec{k})=\sum_i f_i(\vec{k}) e^{2i\pi\vec{k}\cdot\vec{r_i}}
\end{equation}
where $f_i(\vec{k})$ and $\vec{r_i}$ are respectively the scattering factor and the position of atom $i$. In the case of a strained structure, the position of atom $i$ can be written as:
\begin{equation}
\vec{r_i}=\vec{r_i^0}+\vec{u_i}
\end{equation}
where $\vec{r_i^0}$ is the ideal position of atom $i$ in the unstrained crystal and $\vec{u_i}$ its displacement from that position. If the crystal is large compared to the size of the unit cell, and if the deformation varies slowly (i.e. all elements of the strain tensor are $\ll1$ anywhere in the crystal), the displacement in the crystal can be described using block unit cells (\footnote{This is only possible if the calculation is made around a symmetry-authorized Bragg reflection. In the case of a forbidden reflection, small relative atomic displacements within the same unit cell, as well as partially filled unit cells, can induce strong variations of the intensity of the 'forbidden' reflection \cite{richard_defect_2007}.}):
\begin{equation}
\vec{r_{i,j}}=\vec{R_j^0}+\vec{U_j}+\vec{r_i}
\end{equation}
where $\vec{r_{i,j}}$ denotes the absolute position of atom $i$ in unit cell $j$, $\vec{r_i}$ is the position of the atom $i$ relative to the unit cell, $\vec{R_j^0}$ and $\vec{U_j}$ are the ideal position and the displacement vector of unit cell $j$. The scattering can then be rewritten as:
\begin{equation}\label{eqn:eqn5} 
\tilde{A}(\vec{k})=\sum_i\sum_j f_i(\vec{k})\ e^{2i\pi\vec{k}\cdot(\vec{R_j^0}+\vec{U_j}+\vec{r_i})}
=F(\vec{k}) \sum_j e^{2i\pi\vec{k}\cdot(\vec{R_j^0}+\vec{U_j})} 
\end{equation}
where $F(\vec{k})=\sum_i f_i(\vec{k}) e^{2i\pi\vec{k}\cdot\vec{r_i}}$ is the structure factor of the crystal, i.e. this is only valid if all unit cells present the same crystallographic components. If $\Vert(\vec{k}-\vec{k_0})\cdot\vec{U_j}\Vert\ll1$ (where $\vec{k_0}$ is the scattering vector at the position of the Bragg reflection) then we can approximate $\vec{k}\cdot\vec{U_j}$ by $\vec{k_0}\cdot\vec{U_j}$  and we obtain the Fourier transform:
\begin{equation}
\tilde{A}(\vec{k})=F(\vec{k}) \sum_j e^{2i\pi\vec{k}\cdot\vec{R_j^0}} e^{2i\pi\vec{k}\cdot\vec{U_j}}
\approx F(\vec{k}) FT\left[e^{2i\pi\vec{k_0}\cdot\vec{U_j}}\right]
\end{equation}

\begin{figure}
\begin{center}
\includegraphics[width=160mm]{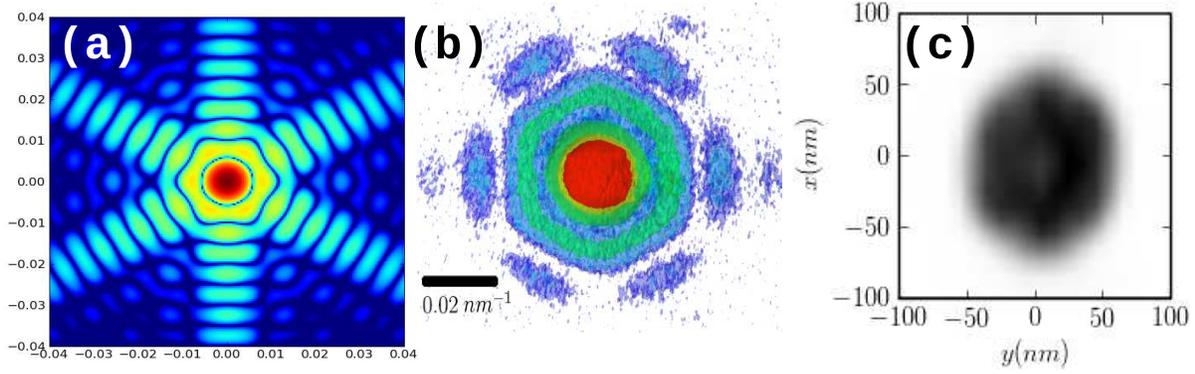}
\end{center}
\caption{\label{fig:fignanowire-silicon} 
Coherent diffraction imaging of a single Si nanowire: (a) simulated 2D scattering pattern (in the plane perpendicular to the nanowire axis), (b) experimental pattern recorded on a 95 nm silicon nanowire \cite{favre-nicolin_coherent-diffraction_2009}, and (c) the corresponding real-space reconstruction of the nanowire cross-section}
\end{figure}

This last approximation is generally used for the analysis of CBI experiments \cite{livet_diffraction_2007,robinson_coherent_2009,chamard_evidence_2008,labat_local_2007} as it allows a fast Fourier transform computation. However, its validity strongly depends on the size of the object and the amplitude of its deformation: e.g. for an object of size 100 unit cells, the extent in reciprocal space of its scattering will be $N*\frac{1}{100}$, where $N$ is the number of oscillations recorded. For $N=10$, and if the maximum deformation on the sides of the crystal is equal to one unit cell (corresponding to a $1\%$ strain over 100 unit cells) then the maximum difference between $\vec{k}\cdot\vec{U_j}$ and $\vec{k_0}\cdot\vec{U_j}$ will be equal to $0.1$, which cannot be neglected ($2\pi\cdot 0.1\approx0.63\ rad \approx 35^\circ$). In such a case it is more accurate to use directly equation \eref{eqn:eqn2} (if individual atomic positions are known) or equation \eref{eqn:eqn5} (if the displacement field is described per block unit cell), although the calculations are significantly slower.

Finally, it is important to note that the strain sensitivity of a single CBI measurements is limited to the direction parallel to the scattering vector: a successful object reconstruction will yield $2\pi\vec{k}\cdot\vec{U}$ as the phase of the reconstructed density. A full reconstruction of an object deformation will require data collection around three linearly independent reflection scattering vectors \textit{for the same object} \cite{leake_longitudinal_2009,beitra_SRI_2009}, less if symmetry can be used.

\section{Coherent Bragg imaging on single nanowires in the absence of faults}\label{section3} 

\subsection{Homogeneous nanowires}\label{section3.1HomgNW} 
Homogeneous nanowires without any external stress only present strain on the atomic layers that are closest to the surface \cite{huang_coordination-dependent_2008}, so given the limited real-space resolution obtained by CBI (usually at least 5 nm), this outer layer contraction should be negligible. In that case coherent scattering around a Bragg reflection will yield the same signal as for the small angle range, i.e. the Fourier transform of the shape of the nanowire. As the nanowire length is much larger than its diameter, the signal will then be a quasi-2D signal which corresponds to the 2D FT of the nanowire cross-section. Example of simulated and experimental CBI signal recorded on a single silicon nanowire is presented in figure \ref{fig:fignanowire-silicon} (see \cite{favre-nicolin_coherent-diffraction_2009} for details).

As in all X-ray experiments, the phase of the scattered amplitude is lost during the experiment, and must be recovered either using ab initio phase retrieval algorithms \cite{gerchberg_practical_1972,fienup_phase_1982,marchesini_x-ray_2003,wu_reconstruction_2005} or using a model with some \textit{a priori} (e.g. symmetry) information \cite{favre-nicolin_coherent-diffraction_2009}.

\subsection{Heterogeneous nanowires}\label{section3.2HeteroNW}
\begin{figure}
\includegraphics[width=160mm]{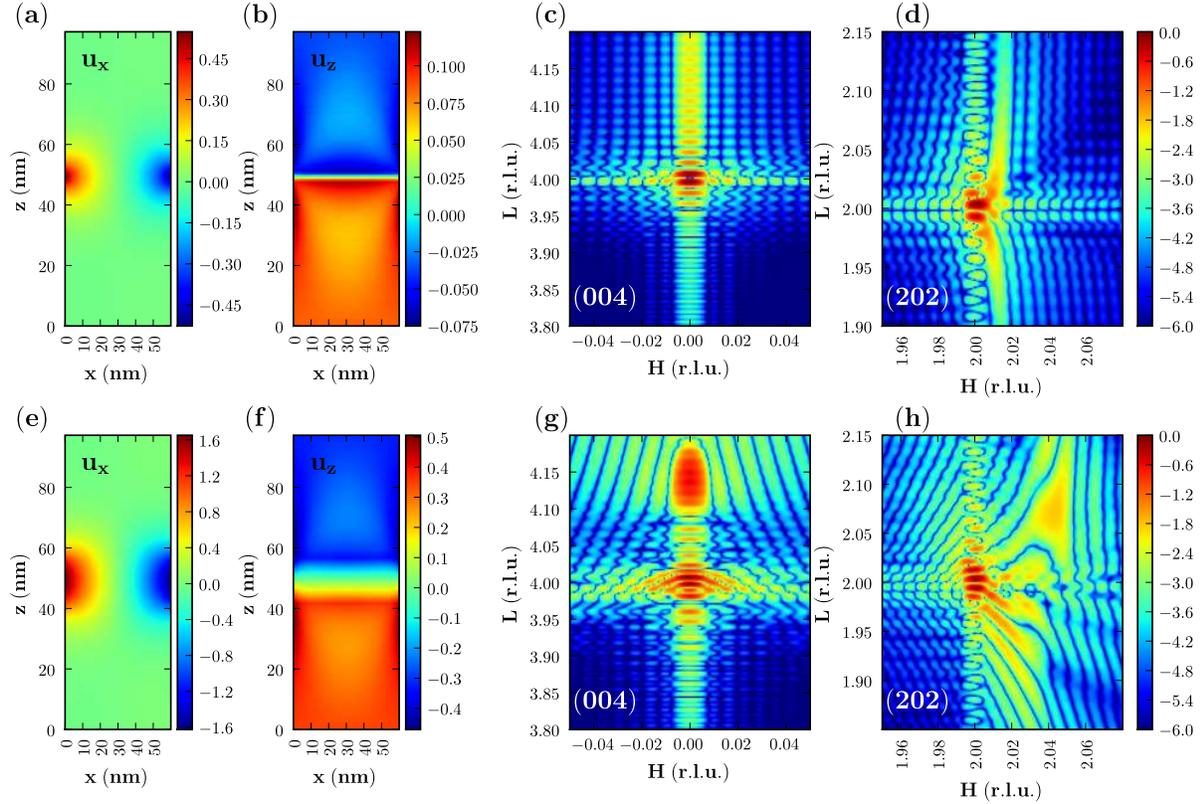}
\caption{\label{fig:figInAs-InP-simu} 
Atomistic simulations of an InAs wurtzite nanowire ($\oslash=\ 60$ nm, height= $100$ nm) with (a,b,c,d) a 3 nm InP insertion and (e,f,g,h) a 15 nm insertion: (a,e) the radial displacement (along $[110]$, expressed relatively to the perfect InAs lattice, in unit cells (u.c.)), (b,f) the axial displacement (along $[001]$), the calculated intensities around the (c,g) $(004)$ and (d,h) $(202)$ reflections, with a logarithmic colour scale. The map coordinates are given in reciprocal lattice units (r.l.u.) relative to InAs.}
\end{figure}

In the case of an heterogeneous nanowire, CBI will be sensitive to the displacement and the nature of all atoms. We have simulated an $[001]_{WZ}$-oriented InAs nanowire, with an InP insertion (see figure \ref{fig:figInAs-InP-simu}), using atomistic simulations - both sections have a wurtzite structure. The radius of the regular hexagon is R=60 nm, and several wire lengths and insertion thicknesses were simulated (see figure \ref{fig:figInAs-InP-simu}) to test the sensitivity of CBI. The relaxed atomic positions are computed using Keating's valence force field (VFF model) \cite{keating_effect_1966,eymery_strain_2007}.

The amplitude of the displacement field (computed relatively to a perfect InAs lattice) has a maximum value of 1-2 unit cells in the horizontal direction ($u_x$), depending on the thickness of the insertion; in the vertical direction ($u_z$) the maximum displacement varies with the InP thickness $t$, ranging from 0.16 ($t$=1.5 nm) to 0.6 unit cells ($t$=15 nm). The InP insertion has a lower bulk lattice parameter than InAs ($a_{WZ}^{InP}=0.415$ nm, $a_{WZ}^{InAs}=0.4268$ nm), so that the insertion contracts the lattice, as can be seen in figure \ref{fig:figInAs-InP-simu}.

Due to the smaller lattice parameter, scattering from the InP insertion occurs at higher angles: although a clear perturbation in the scattering with t=3 nm can be seen (figure \ref{fig:figInAs-InP-simu}(c) and (d)) - an homogeneous nanowire would present roughly (\footnote{The only difference being due to the amplitude of the atomic scattering factors}) the same signal around the two reflections-, it is only with a larger insertion (figure \ref{fig:figInAs-InP-simu}(g) and (h)) that a well-separated peak appears for the InP insertion, roughly two orders of magnitude less intense than the main InAs peak.

These simulations indicate that it is possible to measure a scattering signal specific to a small (a few nm) insertion, assuming at least 3 or 4 orders of magnitude in the experimental data, which is usually achieved. However, this simulation assumes a \textit{perfect} nanowire, without \textit{any} structural fault, which is difficult to achieve. It is therefore important to evaluate the influence of faults on the coherent scattering, and whether it can hide the scattering signal due to a strained lattice.

\section{Coherent Bragg Imaging on nanowires with stacking faults}\label{section4CBIFaultedNW}
\subsection{Stacking faults in nanowires}\label{section4.1SFinNW}
Stacking faults commonly occur in nanowires \cite{li_size-dependent_2004,persson_solid-phase_2004,verheijen_growth_2006,bao_optical_2008,jeppsson_gaas/gasb_2008,tomioka_control_2008,caroff_p._controlled_2009, ding_structures_2009} and sometimes lead to a periodic twinning structure  \cite{hao_periodically_2006,johansson_structural_2006,korgel_semiconductor_2006,xiong_coherent_2006,algra_twinning_2008}. This is especially true for III-V nanowires, where the nanowire geometry and growth kinetics may stabilize the metastable wurtzite structure instead of the zincblende bulk phase \cite{glas_wurtzite_2007,dubrovskii_growth_2008,dubrovskii_growth_2008-1}.
\\

This section presents first a theoretical study on the influence of stacking faults when studying a faulted nanowire, and then shows how it is experimentally possible to avoid the effect of the stacking faults when studying strain in a nanowire. Finally we show how it is possible to use CBI in order to retrieve statistical information about a fault sequence in a nanowire.

\begin{figure}
\includegraphics[width=160mm]{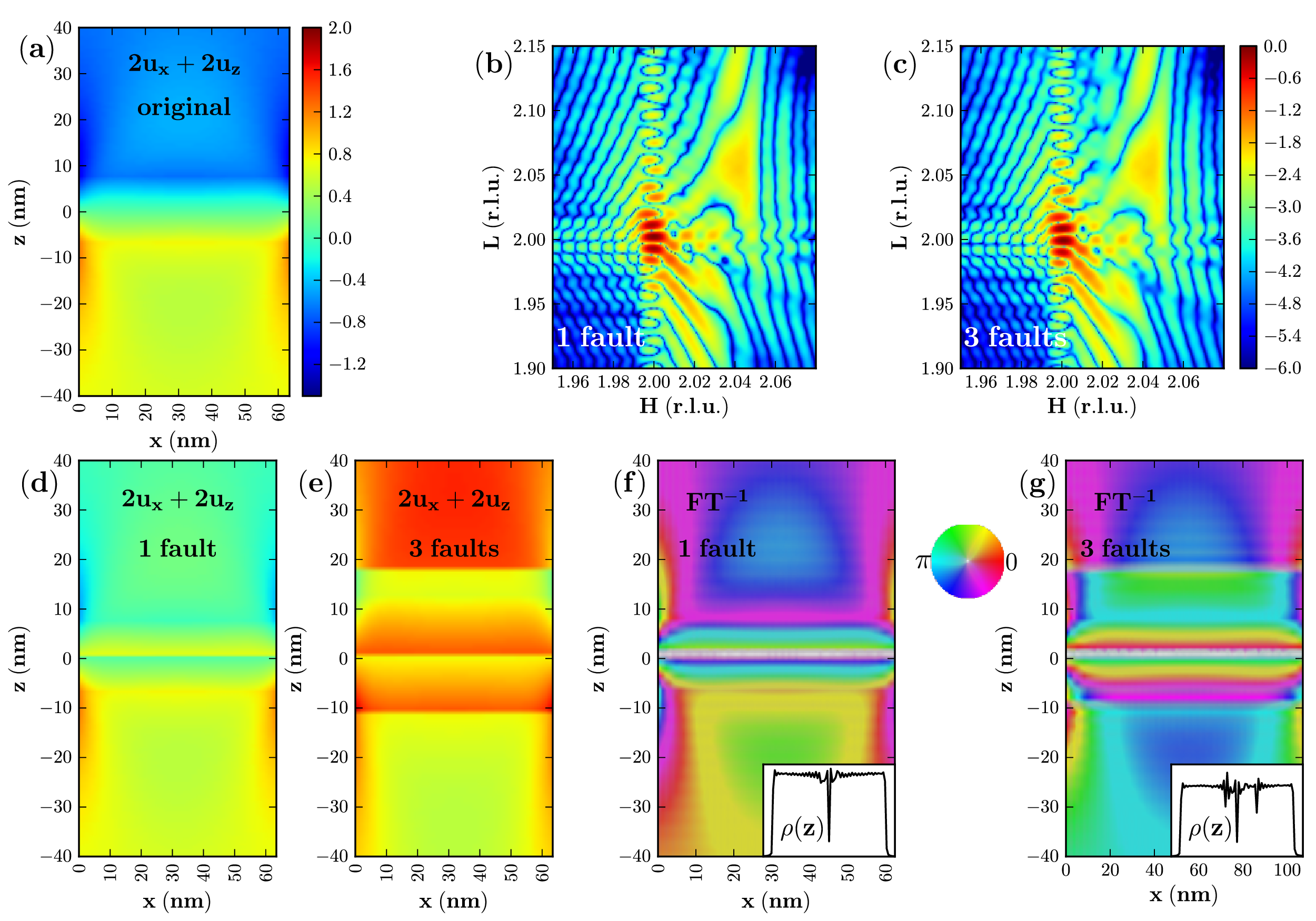}
\caption{\label{fig:fig3} 
Simulation of the influence of stacking faults on the coherent scattering from InAs/InP nanowires ($\oslash=\ 60$ nm, height= $100$ nm) around a $(202)_{WZ}$ reflection. $2u_x+2u_z$ for (a) the original InAs/InP nanowire obtained using atomistic simulations, and in the case of (d) one and (e) three growth faults. The color scale is expressed in unit cells (u.c.) and is the same for (a,d,e). The resulting scattered intensity is shown in (b) (one fault) and (c) (three faults); these can be compared to figure \ref{fig:figInAs-InP-simu}(h) (no fault). The map coordinates are given in reciprocal lattice units (r.l.u.) relative to InAs. There are a number of small changes from the scattering from the original InAs/InP nanowire (see figure \ref{fig:figInAs-InP-simu}(h)), notably with shifted fringes in the domain $H\ \epsilon[2.00;2.04],\ L\ \epsilon[1.95,3.10]$.
The effect of the faults on the reconstruction is more visible when computing the inverse Fourier transform of the simulated complex amplitude around the $(202)$ reflection: the complex recovered density (see text for details) is shown in the case of one (f) and three (g) faults. In these images the phase ($2\pi\vec{k}\cdot\vec{U}$) is given by the colour (as indicated by the colour wheel), while the amplitude is given by the saturation of the colour (the saturation is given on a linear scale: white corresponds to no electronic density and full saturation to the normal average zincblende density). The reconstruction yields in both cases a low density (white) at the position of the structural faults. The inserts show a plot of the amplitude along the vertical axis in the center of the reconstructed wire, where the dips in intensity can be clearly seen.}
\end{figure}

\subsection{CBI in the presence of stacking faults: simulations with InAs/InP nanowires}\label{section4.2SimuFaultsInAs}
During a coherent diffraction experiment, the scattering around a Bragg peak is sensitive to any \textit{displacement} from a perfect periodic lattice. For example in the case of a wurtzite nanowire, a simple deformation fault \cite{krishna_random_1994} corresponds to the following sequence: AB\textbar AB\textbar AB\textbar AB\textbar CA\textbar CA\textbar CA\textbar CA\textbar CA, with the first part being the normal 'AB' sequence, followed by an 'AC' sequence, where all atoms atoms are shifted by $\vec{v}=\frac{1}{3}(1\overline{1}0)_{WZ}$ from their position in the normal sequence. This abrupt shift will produce a specific interference pattern around reflections for which the scattering vector is not orthogonal to $\vec{v}$, and these interferences will cumulate with those coming from the strain-induced deformation.

An example of simulated scattering from InAs/InP nanowires with one and three stacking faults is given in figure \ref{fig:fig3}: in both cases the oscillation fringes are affected by the faults. In theory, such faults would not prevent the reconstruction of the density and displacement fields in the wire by ab initio phase reconstruction algorithms, as the faults correspond to a simple 'phase shift' of $e^{2i\pi\vec{k}\cdot\vec{U}}$ (e.g. for a $\vec{k}=(202)$ reflection and $\vec{v}=\frac{1}{3}(\vec{a}-\vec{b})$, the phase shift is $2\pi\vec{k}\cdot\vec{v}=\frac{4\pi}{3}$).

However in practice, the reconstruction would be hindered by the fact that the resolution of the experiments is currently limited to 5-20 nm (e.g. $\approx$ 10-40 unit cells in the case of InAs), and therefore a displacement configuration such as seen in figure \ref{fig:fig3}(b) may be missed. Moreover, reconstruction algorithms for strained objects tend to converge more slowly than in the unstrained case, as the real-space constraint in the reconstruction can only use the finite support for the object and not its positivity ; for this reason it can be useful to impose continuity to the phase field of the reconstructed object \cite{minkevich_inversion_2007}, a strategy that should be modified to take stacking faults into account.

Moreover, as already pointed out in \cite{chamard_evidence_2008}, the scattering around a given reflection will depend on the domains (zincblende or wurtzite) that are found along the wire: each type of domain may only contribute to the scattering of reflections allowed by its structure, as will be detailed in section \ref{sec:GaAsGaP}.
\\

Finally to quantitatively evaluate the influence of the stacking faults of the nanowire reconstruction from CBI, it is interesting to use the simulated data (figures \ref{fig:fig3} (b) and (c)) and compute the inverse Fourier transform (\footnote{The following method was used: the direct (real space to reciprocal) calculation used was performed using atomic positions as denoted in equation \ref{eqn:eqn2}. The points in reciprocal space at which the scattering was computed were located at the following reciprocal lattice unit coordinates: $h\ \epsilon[1.76,2.4[$, $k\ \epsilon[-.24,.24[$, $l\ \epsilon[1.76,2.4[$, with a step equal to $.004$. The reconstructions (reciprocal to real space) where computed using a Fast Fourier Transform, and the resulting array was cropped to keep only the area containing the crystal - only a 2D section is presented in figure \ref{fig:fig3}.}) to obtain $\rho e^{2i\pi \vec{k}\cdot\vec{U}}$. In order to correctly recover one $(x0z)$ plane in direct space, this was performed from 3D simulated data around the $(202)_{WZ} $ reflections presented in figure \ref{fig:fig3}. The result is presented in figure \ref{fig:fig3}(f) and (g), in the case of one (f) and three (g) inserted growth faults. The deformation field is recovered as the phase of the signal (corresponding to $2\pi(2u_x+2u_z)$), but the most striking features are the 'holes' in the reconstructed amplitude at the position of the faults.

These holes are an artefact caused by the presence of the faults, and can be simply explained: they introduce an (ABC) succession of layers in the middle of the (AB\textbar AB\textbar AB) sequence, and this particular sequence leads to a null intensity. If the first layer (A) is used as a reference, the second (B) is shifted by $\vec{v}_B=\vec{v}+(00\frac{1}{2})=(\frac{1}{3}\frac{\overline{1}}{3}\frac{1}{2})_{WZ}$ and the third by $\vec{v}_C=2\vec{v}+(001)=(\frac{2}{3}\frac{\overline{2}}{3}1)_{WZ}$. Their scattered amplitude around the $(202)_{WZ}$ reflection is then equal to:

\begin{equation}
 1+e^{2i\pi\vec{k}\cdot\vec{v}_B}+e^{2i\pi\vec{k}\cdot\vec{v}_C}=1+e^{\frac{4i\pi}{3}}+e^{\frac{8i\pi}{3}}=0
\end{equation}

As a consequence, this succession of layers of the nanowire contribute no intensity in this part of the reciprocal space, and it logically follows that the inverse Fourier transform from the scattered amplitude around the $(202)$ reflection yields a null electronic density near the faulted parts of the wire.

Recovering the correct intensity near the faults would require a direct space resolution sufficient to resolve individual layers, i.e. half a wurtzite unit cell along the $(001)$ direction. This corresponds to a range in reciprocal space such that $\Delta L>2$, i.e. at least including the $(201)$ and $(203)$ reflections. In practice, it will be much easier to select a reflection such that $\vec{k}\cdot\vec{v}$ is integer and therefore will be insensitive to the faults. An example of different reconstructions using several reflections from a faulted structure can be seen in \cite{beitra_SRI_2009}.

\subsection{CBI in the presence of stacking faults: measurement of one InSb/InP nanowire}\label{sec:InSbInP} 

\begin{figure}
\includegraphics[width=160mm]{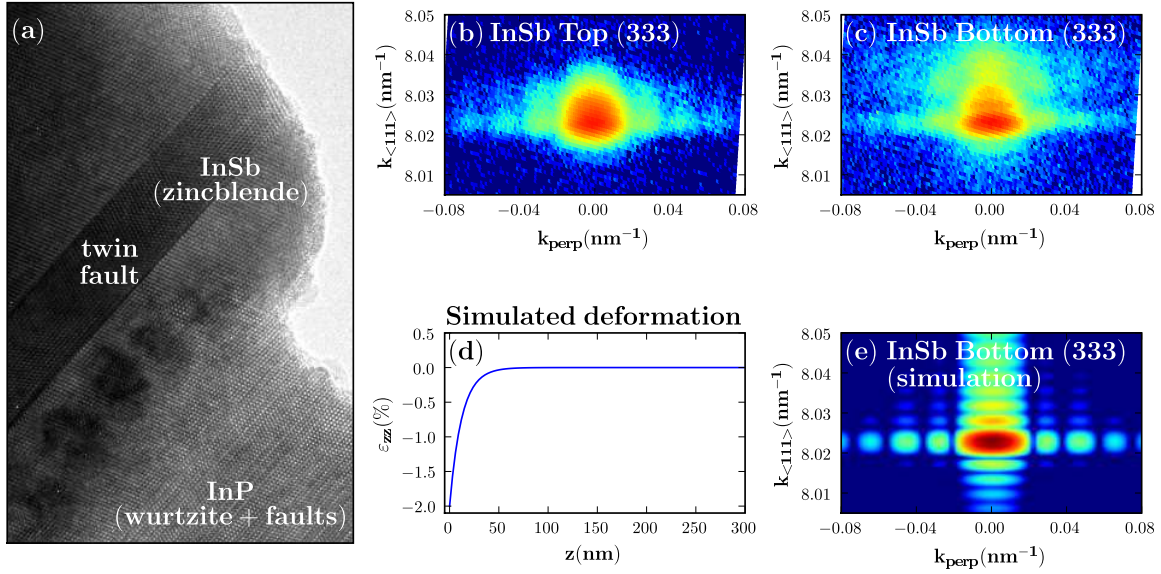}
\caption{\label{fig:fig4} 
(a) Transmission electron microscopy image of an InSb/InP interface, with a twinning fault in the bottom of the InSb part.
Experimental diffraction image ($(3\overline{3}3)_{ZB}$ reflection) from an InSb/InP nanowire, measured at the (b) top and (c) bottom (near the InSb/InP interface) of the InSb section. The lattice parameter of the InP section is sufficiently different ($10.3\%$) not to diffract on the same image.
(d) Simulated deformation as a function of the height near the base of the InSb section (height=250 nm), and (e) the corresponding scattering simulation, which matches well the measurement recorded at the bottom of the wire.
}
\end{figure}

In order to measure the strain from nanowires in the presence of known stacking faults, it is also possible to select a reflection which is insensitive to the displacements associated to the faults, e.g. for a displacement vector 
$\vec{v}$, any Bragg reflection for which $\vec{k}\cdot\vec{v}$ is an integer will be unaffected by the fault.

We tested this approach on InSb/InP heterogeneous nanowires. These nanowires were grown \cite{caroff_InAsInSb_2008,caroff_insb_2009,borg_movpe_2009} from Au seed particles with 30 nm diameter on InP(111)$_B$ substrates using metal-organic vapour phase epitaxy. The InP segment was first grown at 420$^o$C after which the InSb segment was grown at 450$^o$C for 13 min using a V/III ratio of 70. As can be seen in figure \ref{fig:fig4}(a), in these wires the InP section has a wurtzite structure with many stacking faults - this part was not studied to avoid diffraction from the InP substrate. The InSb section has a zincblende structure, with a few (one to three) twinning faults (\ref{fig:fig4}a) located near the base, were there is also a strain relaxation in the first 10-20 nm.

We measured the scattering from single nanowires on the ID01 beamline of the ESRF, using a Fresnel-Zone-Plate (FZP) to focus the 8 keV beam down to $300\ (vert.)\times500\ (horiz.)\ nm^2$ (FWHM), which allowed to select different sections of the nanowire for diffraction (the length of the InSb section is $0.68\pm0.15\ \mu m$). As shown in figure \ref{fig:fig4}(b) and (c) we measured coherent diffraction from both the top and the bottom of the InSb (the sample was vertically translated by 250 nm between the two images). The image corresponding to the bottom of InSb features additional fringes, which cannot be due to twin faults as the chosen reflection -$(333)_{ZB}$- is insensitive to stacking faults. Note that the 'top' image (figure \ref{fig:fig4}(b)) also features an asymmetric shape with some diffuse scattering at larger $k_{<111>}$ values, but that is probably due to some contribution from the base of the wire due to the tails of the focused X-ray beam.

In order to reproduce these asymmetric fringes, we simulated the bottom of the InSb nanowire (height=250 nm, width=54 nm), where the spacing between successive layers (see figure \ref{fig:fig4}(d)) is increased by $\varepsilon_{zz}^0=2\%$ at the bottom, with a simple exponential decrease $\varepsilon_{zz}=\varepsilon_{zz}^0 e^{-z/\Delta}$ ; as the precise relaxation law could not be predicted (the relaxation due to the $10.3\%$ lattice mismatch occurs mainly through dislocations and radial relaxation), we tried several values for $\Delta$, with $\Delta=10$ nm giving the best agreement with the experimental pattern.

The simulated scattering (figure \ref{fig:fig4}(e)) reproduces well the fringes, which are due to the combination of the height of the wire selected by the beam (the fringes period) and the deformation (the asymmetric intensity). The shorter $d_{111}$-spacing at the bottom of InSb (relatively to bulk InSb) is probably due to interdiffusion during the deposition, with an $InSb_{x}P_{1-x}$ chemical composition varying in the first 10-20 nm of InSb (see figure 4(c) in \cite{borg_movpe_2009}).

While this experiment has shown that it is possible to be sensitive to strain near an heterogeneous interface (even in the presence of faults), a new measurement with more intensity and extent in reciprocal space would be necessary to enable the \textit{ab initio} reconstruction of the deformation.

\subsection{CBI with many stacking faults: measurement of a GaAs/GaP nanowire } \label{sec:GaAsGaP} 
\subsubsection{Experimental measurement:}

In the case of a nanowire with many stacking faults, Bragg coherent diffraction can be used to get information about the fault sequence. We studied GaAs/GaP nanowires on beamline 34ID-C of the Advanced Photon Source; during this experiment we used a $1.1\times2.2\ \mu m^2$ beam, obtained with Kirkpatrick-Baez mirrors \cite{kirkpatrick_formation_1948}. Slits were closed down to $20\times50\ \mu m^2$ before the KB mirrors to select the coherent part of the beam. The scattered intensity was collected with a direct illumination CCD camera with $22\ \mu m$ pixel size, placed at 1300 mm from the sample.

The samples (see \cite{verheijen_growth_2006} for synthesis details) consisted of nanowires divided in two parts, with GaP on the bottom and GaAs at the top, and an average diameter of 50 nm (figure \ref{fig:fig5}(a)). They were grown on SiO$_2$ and were randomly oriented. In order to collect the diffraction from the GaAs section (presenting fewer faults), the detector was placed at the $2\theta$ angle corresponding to the $(111)$ reflection, and then the samples were scanned until a diffraction pattern appeared on the detector.

For some wires the recorded images did not present the expected Fourier transform of the nanowire cross-section (figure \ref{fig:fignanowire-silicon}(a)), but rather presented a 'bar-code' pattern, as presented in figure \ref{fig:fig5}(b): from this image the width of the nanowire can be estimated to 50 nm from the horizontal width of the bar-code pattern, while the width along the length (vertical) of the barcode pattern indicates that it is related to domains along the wire which are about 10 nm long.

These measurements performed on a single NW can be compared to those obtained using non-coherent X-ray scattering on an assembly of epitaxial wires: as an example, figure \ref{fig:fig5}(e) shows the experimental intensity recorded along $(20L)_{WZ}$ crystal truncation rods (CTR), on the same InSb/InP nanowires as described in section \ref{sec:InSbInP} and figure \ref{fig:fig4}(a). The grazing-incidence diffraction was recorded on the BM32 beamline of the ESRF, with a $0.05^\circ$ incident angle to minimize the contribution from the InP substrate. The two CTR are extremely different, due to the fact that the InSb part (black line) has a zincblende structure with few faults (\footnote{There are two twin variants of the zincblende InSb, but as one wire is of a given variant (with up to to three twin faults), it only contributes to one CTR and therefore does not enlarge the Bragg peak. See figure 3 in \cite{eymery_strain_2007} for a description of the scattering of the different variants.}) (e.g. the sharp peak around $(20\frac{2}{3})_{WZ}=(311)_{ZB}$), while the InP section (red curve) is predominently wurtzite with many defects, and only features much larger peaks. Note that this type of CTR recorded on an assembly of nanowires can be analysed using statistical models of the stacking sequence \cite{krishna_random_1994}.

\subsubsection{Model for the faulted nanowires:}

In the case of $<111>$ nanowires with stacking faults, and for the $(\overline{1}11)_{ZB}=(10\frac{2}{3})_{WZ}$  reflection (\footnote{Using as basis vectors in direct space: $\vec{a}_{WZ}=\frac{1}{2}(\overline{1}10)_{ZB}$, $\vec{b}_{WZ}=\frac{1}{2}(0\overline{1}1)_{ZB}$ and $\vec{c}_{WZ}=(111)_{ZB}$}) , the phase shift of the scattered signal is $e^{2i\pi \vec{k}\cdot\vec{v}}$, where $\vec{v_{WZ}}=\pm\frac{1}{3}(\vec{a}-\vec{b})$. Depending on the shift from an arbitrary reference stacking, each domain along the wire then diffracts with a phase shift $\varphi \in [-\frac{2\pi}{3},0,+\frac{2\pi}{3}]$, i.e. the nanowire is seen as a 'phase object', and the scattered signal is equal to the Fourier Transform of $e^{i\varphi}$.

The configuration along the wire can be relatively complex: while GaAs as a bulk crystal is stable in the zincblende structure, GaAs nanowires can either be wurtzite \cite{glas_wurtzite_2007} or zincblende (see figure 2(c) in \cite{verheijen_growth_2006}). They can therefore be expected to have 4 types of domains, with two possible twins for zincblende and wurtzite, and each of these domains can be shifted by $\vec{v}$ or $2\vec{v}$ from a reference configuration. The scattering from these different domains is however relatively simple, as only one zincblende twin configuration will contribute to the scattering: the $(\overline{1}11)_{ZB}$ reflection corresponds to the $(10\frac{2}{3})$ in wurtzite reciprocal lattice coordinates, and therefore has no intensity. As for the twin zincblende structure (which is the image of the normal zincblende with respect to a mirror perpendicular to the $<111>$ direction), the corresponding reciprocal lattice coordinates are $\frac{1}{3}(\overline{5}11)$, which does not exist either. As a consequence the reconstructed domain structure from the CDI data should yield both the type of domain, diffracting (normal zincblende) or not (twin zincblende or wurtzite) and the corresponding phase shift of these domains.

\subsubsection{Domain reconstruction:}

\begin{figure}
\includegraphics[width=160mm]{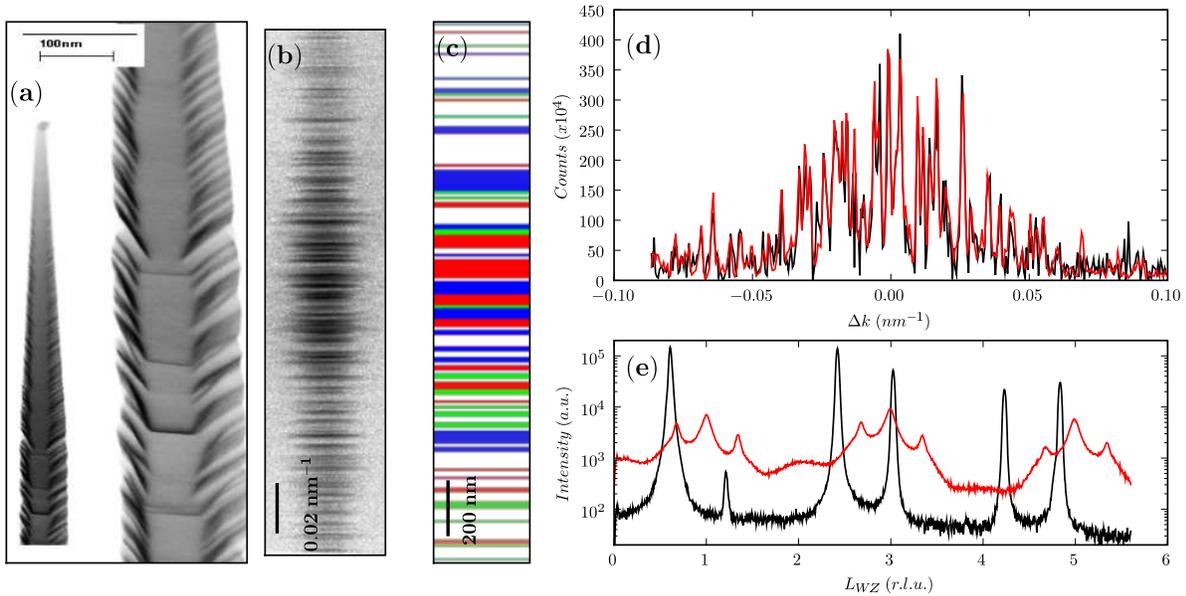}
\caption{\label{fig:fig5}
(a) Scanning electron microscopy image of a $<111>_{ZB}$ nanowire with a GaP bottom section and a top GaAs section. Note the nano-facetting of the wire, which can be related to stacking faults along the length of the wire.
(b) Coherent diffraction image recorded on a $(\overline{1}11)_{ZB}$ GaAs reflection. The "barcode" pattern is due to the faults: the nanowire can be modelled as a stacking of domains with $\pm\frac{2\pi}{3}$ phase shifts (see text for details).
(c) Reconstructed phase object from the diffraction data: each domain along the wire corresponds to a phase $\varphi \in [-\frac{2\pi}{3},0,+\frac{2\pi}{3}]$- associated with colors red,green and blue, or white for a domain not contributing to the scattering of the $(\overline{1}11)_{ZB}$ reflection. The intensity of the color corresponds to the amplitude of the beam (gaussian with FWHM=$1.1\ \mu m$) along the wire.
(d) Integrated 1D profile of the 'barcode' pattern, observed (red) and simulated (black) using the phase object in (c).
(e) Non-coherent crystal truncation rods (CTR) measured on an assembly of InSb/InP nanowires: the CTR from InSb (black), which has a zincblende structure with very few faults, presents very sharp peaks, whereas the InP CTR (red) is very wide, due to the high density of stacking faults in the wurtzite structure.
}
\end{figure}

In order to reconstruct the phase sequence from the scattering data, it is theoretically possible to retrieve the phase from the scattered signal using phase retrieval algorithms (and then using an inverse Fourier Transform to reconstruct the object), but this requires that the data be sufficiently oversampled. In our case, the barcode pattern was very sharp (see figures \ref{fig:fig5}(b) and (d)), with some peaks as small as 2 pixels, and the wire was longer than the beam. As the number of parameters was limited (400 integrated data points), we decided to retrieve the phases of each domain in direct space, using parallel tempering, a biased reverse Monte-Carlo algorithm \cite{falcioni_biased_1999}. During each cycle of this algorithm, either a small number of phases or an occupancy factor (1 or 0) along the wire are randomly changed. The new configuration is then accepted if the R-factor ($R=\sqrt{\frac{\sum_i{(I_{obs}^i-I_{calc}^i)^2}}{\sum_i{(I_{obs}^i})^2}}$) diminishes, and otherwise the configuration is rejected with a probability $P=e^{-(R-R_{previous})/T}$, where T is the temperature of the algorithm.

The object itself was modelled as a 1D object with $\approx5$ nm domains, each with a phase $\varphi \in [-\frac{2\pi}{3},0,+\frac{2\pi}{3}]$ and with an occupancy (1 for a diffracting domain, 0 otherwise). As the GaAs part of the wire was longer than the beam at the focal point, the beam was modelled with a gaussian shape, with a $1.1\ \mu m$ beam size. The reconstructed phase object is presented in figure \ref{fig:fig5}(c). Note that the reconstructed object is not unique, which is a known issue in the case of 1D problems \cite{crimmins_ambiguity_1981}. However the analysis of several solutions show that they have the same statistical information (number and size of each type of domain), as the uncertainties given below indicate.

Quantitative information can be obtained from the reconstruction. First, the size distribution of the normal zincblende domains (the only ones seen by the reconstruction) is as follows: 5 nm ($54\pm1\%$), 10 nm ($28\pm1\%$),\ 15\ nm ($8\pm4\%$), 20 nm ($8\pm4\%$), with an average domain size of $9\pm0.5$ nm. This value is comparable to the size that can be estimated from the FWHM of the barcode pattern (figure \ref{fig:fig5}(d)), $L=\frac{1}{.09}=11$ nm. This length is also in agreement with the apparent frequency of 'kinks' on figure \ref{fig:fig5}(a), which are known \cite{johansson_structural_2006,korgel_semiconductor_2006,algra_twinning_2008,caroff_p._controlled_2009} to originate from a succession of zincblende twins.

Furthermore the results of several (10) optimization runs indicate a percentage of occupied (normal zincblende) domains slightly lower than $50\%$: $P(normal\ zincblende)=46\pm3\%$. As the two zincblende twins should be equiprobable (any bias introduced by the terminating planes of the GaP section should vanish after a few domain changes), this indicates that the GaAs section consists of $92\pm6\%$ zincblende domains, and therefore of $8\pm6\%$ wurtzite domains.
\\
\subsubsection{Discussion on improvement to the \textit{ab initio} domain reconstruction:}

Structures like the `bar code' pattern are found to be particularly difficult to phase and this has not yet been possible with traditional methods \cite{marchesini_coherent_2003,fienup_phase_1982}.  There are several possible reasons for this: first and foremost is that the interference pattern is only one-dimensional (1D), even though the full recorded pattern is 2D.  As mentioned above, the cross-section is just the shape transform of the nanowire itself; without the faults the diffraction pattern would be a single peak, being a disk viewed edge-on.  1D diffraction patterns are prone to factorisation of their amplitude functions, so the object can be expressed as a convolution of substructures; 2D and 3D functions are much less likely to factorise \cite{crimmins_ambiguity_1981}.  The complex conjugate of each of these pieces gives the same modulus for its factor, so there are a large number of equivalent solutions \cite{crimmins_ambiguity_1981}.

The second problem, somewhat related to the first, is that the sequence of nanowire faulted segments tends to be self similar, with internal repeats; permutations of these repeating subsequences give virtually the same diffraction pattern, so the solution is non-unique, especially in the presence of noise.  This non-uniqueness was indeed seen with the biased Monte Carlo method \cite{falcioni_biased_1999} described above, just as was also found with the 2D FeAl antiphase domain inversion results reported previously \cite{stadler_coherent_2007}.

\begin{figure}
\includegraphics[width=160mm]{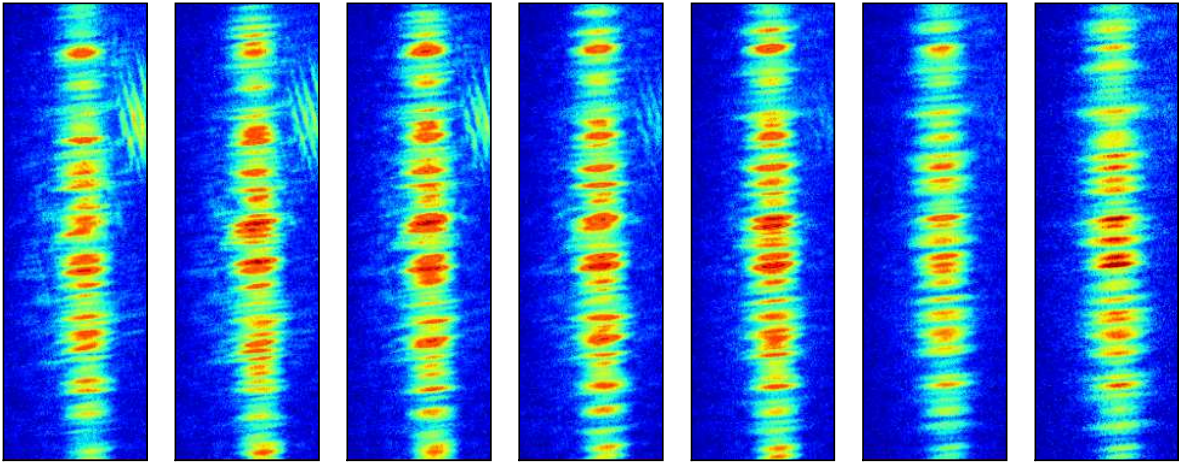}
\caption{\label{fig:fig6}
Ptychography data obtained by moving the focussed beam in 1-micron steps, recorded for the same GaAs/GaP nanowire as shown in figure \ref{fig:fig5}.  The sequence of diffraction patterns shows a steady evolution of the positions of the maxima and minima.
}
\end{figure}

A possible solution to the uniqueness problem has been proposed by Rodenburg {\it et. al.} in a method called ``Ptychography", in which a well-defined beam illumination function is scanned over the region of interest in the sample, along the wire in this case.  The overlapping diffraction patterns that result contain redundant information and the overlap can be made arbitrarily small.  An example of ptychographical data, recorded for the same GaAs/GaP nanowire, is shown in figure \ref{fig:fig6} ; the individual diffraction patterns are clearly related to each other, but show differences that contain additional information about the structure. An algorithm for the recovery of the structure has been proposed \cite{rodenburg_hard-x-ray_2007}, but not yet tried with these data.\\

Partial coherence is another important issue that may limit the effectiveness of the phasing methods described here.  Typically, the size of the objects discussed here is considerably smaller than the lateral coherence length of the X-ray beams used.  Yet partial coherence effects can be seen directly in the graph of figure \ref{fig:fig5}(d): the visibility of the data (red) is less than that of the best-fit simulation (black). The data rarely come down near zero intensity, while the simulation comes closer; this is because there is a second intensity component present in the data, which does not interfere with the main one.  Given that the limited lateral coherence originates from the finite size of the synchrotron X-ray source, it can be modelled as a Gaussian smearing of the resulting diffraction patterns \cite{williams_coherent_2007}, but this does not lend itself to easy methods of removing it.

A more interesting way forward is offered by the new work of Flewett {\it et. al.} \cite{flewett_extracting_2009}, who show that a typical synchrotron source can be modelled by a small but finite number of modes, with  more than $90\%$ in the primary mode in the case studied. The additional modes can be attributed to the edges of the source, particularly in the horizontal direction.  These extra modes, with an illumination profile that is orthogonal to the main mode, give different diffraction patterns from the same object; these are combined (as intensities) in the diffraction pattern which therefore has lower contrast (visibility). So long as the extra modes are a small fraction of the main mode, they can be used to correct for the distortions of the data within an algorithm, without adversely affecting the iterative computation on the main mode. Again, this method has not yet been tried on the data of figure \ref{fig:fig5}.\\

\section{Conclusion}\label{section5Conclusion}
During the last ten years, coherent X-ray scattering has seen very significant progress, notably through the use of better focusing optics, producing a smaller beam size with a much higher photon flux ($5.10^3-5.10^4$ ph/s/nm$^2$). This allowed to study objects smaller than $100$ nm, both for small-angle and Bragg CDI experiments.

The development of Bragg CDI is particularly important as it gives access to the deformation in the volume of objects at the nanoscale, which is essential for the understanding of their structural properties. This point has been illustrated by our experimental and simulation studies on InAs/InP and InSb/InP nanowires. In the case of faulted structures, CBI can be used to retrieve a statistical image (size and type of domains) of the wire. The most important development is probably the ability to use the very small size of the beam to study different parts of a given heterostructure.

However as we have shown in this study, information about the shape and strain state can be hindered by the presence of structural faults, such as stacking faults in nanowires - as we have shown this can lead to voids in the reconstructed electronic density. While CBI allows a quantitative analysis of the structural faults, illustrated here in the GaAs system, the combined presence of strain and stacking faults in a structure remains a complicated problem - in the simplest case (stacking faults) the use of carefully selected reflections will allow to be insensitive to the stacking order, but this requires the ability to study several reflections for the same object, which is still difficult for objects smaller than 100 nm, as the mechanical precision (confusion sphere) of existing goniometers is much larger than this value, especially when performing several rotations to select different reflections.

\ack
This work has been partially performed under the EU program NODE 015783. The authors would like to thank the ID01 staff (P. Boesecke, O. Bikondoa, G. Carbone, A. Diaz) for their help during the ESRF experiments, and both the ESRF and the APS for providing beamtime for the experiments.

\section*{References}
\bibliography{article}

\begin{thebibliography}{10}

\bibitem{miao_extendingmethodology_1999}
Miao J, Charalambous P, Kirz J, and Sayre D.
\newblock {\em Nature}, 400(6742):342--344, July 1999.

\bibitem{robinson_reconstruction_2001}
Robinson~I K, Vartanyants~I A, Williams~G J, Pfeifer~M A, and Pitney~J A.
\newblock {\em Physical Review Letters}, 87(19):195505, October 2001.

\bibitem{pfeifer_three-dimensional_2006}
Pfeifer~M A, Williams~G J, Vartanyants~I A, Harder R, and RobinsonI K.
\newblock {\em Nature}, 442(7098):63--66, July 2006.

\bibitem{livet_diffraction_2007}
Livet F.
\newblock {\em Acta Crystallographica A}, 63(2):87--107, 2007.

\bibitem{snigirev_high_2008}
Snigirev A and Snigireva I.
\newblock {\em Comptes Rendus Physique}, 9(5-6):507--516, June 2008.

\bibitem{snigirev_focusing_1998}
Snigirev A, Kohn V, Snigireva I, Souvorov A, and Lengeler B.
\newblock {\em Applied Optics}, 37(4):653--662, February 1998.

\bibitem{kirkpatrick_formation_1948}
Kirkpatrick P and Baez~A V.
\newblock {\em Journal of the Optical Society of America}, 38(9):766--773,
  1948.

\bibitem{baez_fresnel_1961}
Baez~A V.
\newblock {\em Journal of the Optical Society of America}, 51(4):405--412,
  April 1961.

\bibitem{di_fabrizio_high-efficiency_1999}
Di~Fabrizio E, Romanato F, Gentili M, Cabrini S, Kaulich B, Susini J, and
  Barrett R.
\newblock {\em Nature}, 401(6756):895--898, October 1999.

\bibitem{schroer_coherent_2008}
Schroer~C G, Boye P, Feldkamp~J M, Patommel J, Schropp A, Schwab A, Stephan S,
  Burghammer M, Schoder S, and Riekel C.
\newblock {\em Physical Review Letters}, 101(9):090801--4, 2008.

\bibitem{miao_approach_2001}
Miao J, Hodgson~K O, and Sayre D.
\newblock {\em Proceedings of the National Academy of Sciences of the United
  States of America}, 98(12):6641--6645, June 2001.

\bibitem{barty_three-dimensional_2008}
Barty A, Marchesini S, Chapman~H N, Cui C, Howells~M R, Shapiro~D A, Minor~A M,
  Spence J~C H, Weierstall U, Ilavsky J, Noy A, {Hau-Riege}~S P, Artyukhin~A B,
  Baumann T, Willey T, Stolken J, van Buuren~T, and Kinney~J H.
\newblock {\em Physical Review Letters}, 101(5):055501--4, 2008.

\bibitem{marchesini_coherent_2003}
Marchesini S, Chapman H, {Hau-Riege} S, London R, Szoke, He~H, Howells M,
  Padmore H, Rosen R, Spence J, and Weierstall U.
\newblock {\em Optics Express}, 11(19):2344--2353, 2003.

\bibitem{gerchberg_practical_1972}
Gerchberg R Wand Saxton~W O.
\newblock {\em Optik}, 35(2):237--246, 1972.

\bibitem{fienup_phase_1982}
Fienup~J R.
\newblock {\em Applied Optics}, 21(15):2758--2769, 1982.

\bibitem{marchesini_x-ray_2003}
Marchesini S, He~H, Chapman H Nand {Hau-Riege}~S P, Noy A, Howells~M R,
  Weierstall U, and Spence J~C H.
\newblock {\em Physical Review B}, 68(14):140101, October 2003.

\bibitem{wu_reconstruction_2005}
Wu~J S and Spence J~C H.
\newblock {\em Acta Crystallographica A}, 61(2):194--200, 2005.

\bibitem{williams_three-dimensional_2003}
Williams~G J, Pfeifer~M A, Vartanyants~I A, and Robinson~I K.
\newblock {\em Physical Review Letters}, 90(17):175501, April 2003.
\newblock Copyright {(C)} 2009 The American Physical Society; Please report any
  problems to prola@aps.org.

\bibitem{stadler_coherent_2007}
Stadler L-M, Harder R, Robinson~I K, Rentenberger C, Karnthaler H-P, Sepiol B,
  and Vogl G.
\newblock {\em Physical Review B}, 76(1):014204--9, July 2007.

\bibitem{harder_orientation_2007}
Harder R, Pfeifer~M A, Williams~G J, Vartaniants~I A, and Robinson~I K.
\newblock {\em Physical Review B}, 76(11):115425--4, 2007.

\bibitem{robinson_coherent_2009}
Robinson I and Harder R.
\newblock {\em Nature Materials}, 8(4):291--298, April 2009.

\bibitem{chamard_evidence_2008}
Chamard V, Stangl J, Labat S, Mandl B, Lechner~R T, and Metzger~T H.
\newblock {\em Journal of Applied Crystallography}, 41(2):272--280, 2008.

\bibitem{favre-nicolin_coherent-diffraction_2009}
{Favre-Nicolin} V, Eymery J, Koester R, and Gentile P.
\newblock {\em Physical Review B}, 79(19):195401--5, May 2009.

\bibitem{diaz_coherent_2009}
Diaz A, Mocuta C, Stangl J, Mandl B, David C, {Vila-Comamala} J, Chamard V,
  Metzger, and Bauer G.
\newblock {\em Physical Review B}, 79(12):125324--5, March 2009.

\bibitem{leake_longitudinal_2009}
Leake~S J, Newton~M C, Harder R, and Robinson~I K.
\newblock {\em Optics Express}, 17(18):15853--15859, 2009.

\bibitem{mocuta_beyondensemble_2008}
Mocuta C, Stangl J, Mundboth K, Metzger~T H, Bauer G, Vartanyants~I A,
  Schmidbauer M, and Boeck T.
\newblock {\em Physical Review B}, 77(24):245425--6, June 2008.

\bibitem{scheler_probingelastic_2009}
Scheler T, Rodrigues M, Cornelius~T W, Mocuta C, Malachias A,
  {Magalhaes-Paniago} R, Comin F, Chevrier J, and Metzger~T H.
\newblock {\em Applied Physics Letters}, 94(2):023109--3, 2009.

\bibitem{diaz_spatially_2009}
Diaz A, Mocuta C, Stangl J, {Vila-Comamala} J, David C, Metzger~T H, and Bauer
  G.
\newblock {\em Physica Status Solidi A (Applications and Materials Science)},
  206(8):1829--1832, August 2009.

\bibitem{krishna_random_1994}
Krishna P and Sebastian~M T.
\newblock Gordon \& Breach Science Publishers Ltd, illustrated edition edition,
  March 1994.

\bibitem{richard_defect_2007}
Richard M-I, Metzger~T H, Holy V, and Nordlund K.
\newblock {\em Physical Review Letters}, 99(22):225504--4, November 2007.

\bibitem{labat_local_2007}
Labat S, Chamard V, and Thomas O.
\newblock {\em Thin Solid Films}, 515(14):5557--5562, May 2007.

\bibitem{beitra_SRI_2009}
Beitra L, Watari M, Matsuura T, Shimamoto N, Harder R, and Robinson I.
\newblock In {\em Submitted to: proceeedings of the International Conference on
  Synchrotron Radiation and Instrumentation, 2009. {SRI} '09}.

\bibitem{huang_coordination-dependent_2008}
Huang~W J, Sun R, Tao J, Menard~L D, Nuzzo~R G, and Zuo~J M.
\newblock {\em Nature Materials}, 7(4):308--313, April 2008.

\bibitem{keating_effect_1966}
Keating~P N.
\newblock {\em Physical Review}, 145(2):637, May 1966.

\bibitem{eymery_strain_2007}
Eymery J, Rieutord F, {Favre-Nicolin} V, Robach O, Niquet Y-M, Froberg L, M\aa
  rtensson T, and Samuelson L.
\newblock {\em Nano Letters}, 7(9):2596--2601, 2007.

\bibitem{li_size-dependent_2004}
Li~Q, Gong X, Wang C, Wang J, Ip~K, and Hark S.
\newblock {\em Advanced Materials}, 16(16):1436--1440, 2004.

\bibitem{persson_solid-phase_2004}
Persson~A I, Larsson~M W, Stenstrom S, Ohlsson~B J, Samuelson L, and
  Wallenberg~L R.
\newblock {\em Nature Materials}, 3(10):677--681, October 2004.

\bibitem{verheijen_growth_2006}
Verheijen~M A, Immink G, de~Smet~T, Borgstrom~M T, and Bakkers E P~A M.
\newblock {\em Journal of the American Chemical Society}, 128(4):1353--1359,
  February 2006.

\bibitem{bao_optical_2008}
Bao J, Bell~D C, Capasso F, Wagner~J B, M\aa rtensson T, Tr\"{a}g\aa rdh J, and
  Samuelson L.
\newblock {\em Nano Letters}, 8(3):836--841, March 2008.

\bibitem{jeppsson_gaas/gasb_2008}
Jeppsson M, Dick~K A, Wagner~J B, Caroff P, Deppert K, Samuelson L, and
  Wernersson L-E.
\newblock {\em Journal of Crystal Growth}, 310(18):4115--4121, August 2008.

\bibitem{tomioka_control_2008}
Tomioka K, Motohisa J, Hara S, and Fukui T.
\newblock {\em Nano Letters}, 8(10):3475--3480, October 2008.

\bibitem{caroff_p._controlled_2009}
Caroff P, Dick~K A, {Johansson} J, {Messing}~M E, {Deppert} K, and {Samuelson}
  L.
\newblock {\em Nature Nanotechnology}, 4(1):50--55, 2009.

\bibitem{ding_structures_2009}
Ding Y and Wang~Z L.
\newblock {\em Micron}, 40(3):335--342, April 2009.

\bibitem{hao_periodically_2006}
Hao Y, Meng G, Wang~Z L, Ye~C, and Zhang L.
\newblock {\em Nano Letters}, 6(8):1650--1655, 2006.

\bibitem{johansson_structural_2006}
Johansson J, Karlsson~L S, Svensson C~P T, M\aa rtensson T, Wacaser~B A,
  Deppert K, Samuelson L, and Seifert W.
\newblock {\em Nature Materials}, 5(7):574--580, July 2006.

\bibitem{korgel_semiconductor_2006}
Korgel~B A.
\newblock {\em Nature Materials}, 5(7):521--522, July 2006.

\bibitem{xiong_coherent_2006}
Xiong Q, Wang J, and Eklund~P C.
\newblock {\em Nano Letters}, 6(12):2736--2742, December 2006.

\bibitem{algra_twinning_2008}
Algra~R E, Verheijen~M A, Borgstrom~M T, Feiner L-F, Immink G, van Enckevort W
  J~P, Vlieg E, and Bakkers E P~A M.
\newblock {\em Nature}, 456(7220):369--372, November 2008.

\bibitem{glas_wurtzite_2007}
Glas F, Harmand J-C, and Patriarche G.
\newblock {\em Physical Review Letters}, 99(14):146101--4, October 2007.

\bibitem{dubrovskii_growth_2008}
Dubrovskii~V G, Sibirev~N V, Harmand~J C, and Glas F.
\newblock {\em Physical Review B}, 78(23):235301--10, December 2008.

\bibitem{dubrovskii_growth_2008-1}
Dubrovskii~V G and Sibirev~N V.
\newblock {\em Physical Review B}, 77(3):035414--8, 2008.

\bibitem{minkevich_inversion_2007}
Minkevich~A A, Gailhanou M, Micha J-S, Charlet B, Chamard V, and Thomas O.
\newblock {\em Physical Review B}, 76(10):104106--5, 2007.

\bibitem{caroff_InAsInSb_2008}
Caroff P, Wagner~J B, Dick~K A, Nilsson~H A, Jeppsson M, Deppert K, Samuelson
  L, Wallenberg~L R, and Wernersson L-E.
\newblock {\em Small}, 4(7):878--882, 2008.

\bibitem{caroff_insb_2009}
Caroff P, Messing~M E, Borg~B M, Dick~K A, Deppert K, and Wernersson L-E.
\newblock {\em Nanotechnology}, 20(49):495606, 2009.

\bibitem{borg_movpe_2009}
Borg~B M, Messing~M E, Caroff P, Dick~K A, Deppert K, and Wernersson L-E.
\newblock In {\em Indium Phosphide \& Related Materials, 2009. {IPRM} '09.
  {IEEE} International Conference on}, pages 249--252, 2009.

\bibitem{falcioni_biased_1999}
Falcioni M and Deem~M W.
\newblock {\em The Journal of Chemical Physics}, 110(3):1754--1766, 1999.

\bibitem{crimmins_ambiguity_1981}
Crimmins~T R and Fienup~J R.
\newblock {\em Journal of the Optical Society of America}, 71(8):1026--1028,
  1981.

\bibitem{rodenburg_hard-x-ray_2007}
Rodenburg~J M, Hurst~A C, Cullis~A G, Dobson~B R, Pfeiffer F, Bunk O, David C,
  Jefimovs K, and Johnson I.
\newblock {\em Physical Review Letters}, 98(3):034801--4, 2007.

\bibitem{williams_coherent_2007}
Williams~G J, Quiney~H M, Peele~A G, and Nugent~K A.
\newblock {\em Physical Review B}, 75(10):104102--7, March 2007.

\bibitem{flewett_extracting_2009}
Flewett F, Quiney~H M, Tran~C Q, and Nugent~K A.
\newblock {\em Optics Letters}, 34(14):2198--2200, July 2009.

\end{thebibliography}
\bibliographystyle{unsrt} 

\end{document}